\pdfoutput=1
\documentclass[11pt]{article}
\usepackage{amsmath}

\usepackage[final]{acl}

\usepackage{times}
\usepackage{latexsym}

\usepackage[T1]{fontenc}
\usepackage[utf8]{inputenc}
\usepackage{booktabs}

\usepackage{microtype}
\usepackage{stfloats}

\usepackage{inconsolata}

\usepackage{graphicx}
\usepackage{tabularx} % 需要加载 tabularx 宏包
\usepackage{subcaption}

\title{LLaSE-G1: Incentivizing Generalization Capability for LLaMA-based Speech Enhancement}

\author{
    \textbf{Boyi Kang}\textsuperscript{1} \thanks{Equal contribution.},
    \textbf{Xinfa Zhu}\textsuperscript{1} \footnotemark[1],
    \textbf{Zihan Zhang}\textsuperscript{1} \footnotemark[1], 
    \textbf{Zhen Ye}\textsuperscript{2},
    \textbf{Mingshuai Liu}\textsuperscript{1},
    \textbf{Ziqian Wang}\textsuperscript{1}, \\
    \textbf{Yike Zhu}\textsuperscript{1}, 
    \textbf{Guobin Ma}\textsuperscript{1}, 
    \textbf{Jun Chen}\textsuperscript{3},
    \textbf{Longshuai Xiao}\textsuperscript{3}, 
    \textbf{Chao Weng}\textsuperscript{3}, 
    \textbf{Wei Xue}\textsuperscript{2}, 
    \textbf{Lei Xie}\textsuperscript{1} \\
    \textsuperscript{1} Northwestern Polytechnical University \\
    \textsuperscript{2} The Hong Kong University of Science and Technology \\
    \textsuperscript{3} Huawei Technologies Co., Ltd.\\
    \texttt{\{beaukang, xfzhu, zhzhang, lxie\}@mail.nwpu.edu.cn} \\
}

\begin{document}
\maketitle
\begin{abstract}
Recent advancements in language models (LMs) have demonstrated strong capabilities in semantic understanding and contextual modeling, which have flourished in generative speech enhancement (SE). However, many LM-based SE approaches primarily focus on semantic information, often neglecting the critical role of acoustic information, which leads to acoustic inconsistency after enhancement and limited generalization across diverse SE tasks. In this paper, we introduce LLaSE-G1, a LLaMA-based language model that incentivizes generalization capabilities for speech enhancement. LLaSE-G1 offers the following key contributions: First, to mitigate acoustic inconsistency, LLaSE-G1 employs continuous representations from WavLM as input and predicts speech tokens from X-Codec2, maximizing acoustic preservation. Second, to promote generalization capability, LLaSE-G1 introduces dual-channel inputs and outputs, unifying multiple SE tasks without requiring task-specific IDs. Third, LLaSE-G1 outperforms prior task-specific discriminative and generative SE models, demonstrating scaling effects at test time and emerging capabilities for unseen SE tasks. Additionally, we release our code and models to support further research in this area \footnote{\href{https://github.com/Kevin-naticl/LLaSE-G1}{LLaSE-G1 Codes and Demos}}. %\footnote{\href{https://submission-papers.github.io/LLaSE-G1-demo-page/}{LLaSE-G1 Demo}}  

\end{abstract}

\section{Introduction}
In recent years, large language models (LLMs) have made significant strides in natural language processing (NLP)~\cite{openai2024gpt4technicalreport}, computer vision (CV)~\cite{tschannen2024givtgenerativeinfinitevocabularytransformers,chang2022maskgitmaskedgenerativeimage}, and speech processing~\cite{valle,speechgpt}, driving the rapid development of artificial intelligence technologies. In the NLP domain, LLMs have redefined text generation benchmarks through innovative pre-training and post-training paradigms, particularly excelling in few-shot and zero-shot learning scenarios.
The impact of LLMs extends beyond unimodal textual processing. In CV research, integrating LLMs with visual models has sparked the rise of multimodal learning frameworks, facilitating more efficient processing of tasks such as image comprehension and generation. Similarly, the convergence of modalities is evident in the speech domain, where LLMs have enhanced the naturalness and accuracy of speech interaction systems. These advancements not only highlight the power of LLMs within individual domains but also underscore their potential for multimodal tasks.

% \usepackage{booktabs}
\begin{table}[ht]
\centering
\footnotesize
\begin{tabular}{@{}lll@{}}
\toprule
Task Type & Distortion                                                                   & Reference Signal      \\ \midrule
NS        & Noise, Reverb                                                                & -                     \\ \hline
PLC       & Noise, Packet Loss                                                           & Lossy Label           \\ \hline
TSE       & \begin{tabular}[c]{@{}l@{}}Noise, Reverb, \\ Interfering Speech\end{tabular} & Enrolled Speech     \\ \hline
AEC       & Noise, Reverb, Echo                                                          & Echo Speech \\ \hline
SS        & \begin{tabular}[c]{@{}l@{}}Noise, Reverb, \\ Interfering Speech\end{tabular} & -                     \\ \bottomrule
\end{tabular}
\caption{Subtasks Definition in Speech Enhancement}
\label{tab:Subtasks}
\end{table}

As a fundamental task in the field of speech processing, \textit{speech enhancement (SE)} aims to remove interference from noisy speech and separate and reconstruct clean target speech. Depending on the differences between the interfering speech and the target speech, sub-tasks can be defined as Noise Suppression (NS), Packet Loss Concealment (PLC), Target Speaker Extraction (TSE), Acoustic Echo Cancellation (AEC), Speech Separation (SS), and others, as detailed in Table~\ref{tab:Subtasks}. Neural SE models can generally be categorized into two types: discriminative~\cite{zhao2024frcrnboostingfeaturerepresentation,zhao2024mossformer2combiningtransformerrnnfree} and generative~\cite{wang2024selmspeechenhancementusing}. Deep learning-based discriminative SE models learn a mapping between degraded speech and the corresponding clean speech target. In contrast, generative SE models employ language models or diffusion models to learn the data distribution of the target speech. Notable recent models, including SELM~\cite{wang2024selmspeechenhancementusing}, TSELM~\cite{tang2024tselmtargetspeakerextraction}, and GenSE~\cite{yao2025gensegenerativespeechenhancement}, leverage semantic understanding and contextual modeling capabilities, achieving competitive performance in speech enhancement tasks. While traditional discriminative SE models require carefully designed architectures and task-specific loss functions, generative SE models offer a more flexible framework, enabling better scalability across different SE tasks.

Despite surpassing traditional discriminative models in speech quality, generative SE models still face challenges in acoustic preservation and task generalization. Many generative SE models rely on discrete speech tokens—typically extracted from speech codecs—as inputs to facilitate language modeling. However, as speech is inherently a continuous signal, using discrete tokens, especially semantic tokens, inevitably results in information loss~\cite{yao2025gensegenerativespeechenhancement}, leading to acoustic inconsistencies after enhancement, such as changes in speaker timbre and intonation. Moreover, most generative models are focused on a single task, such as noise suppression, which limits their generalization across different SE tasks. Since SE tasks differ in their input, output, and underlying functions, it remains an open question whether LMs can serve as versatile, multi-task SE models.

In this paper, we argue that, with appropriate design, a single language model can be a powerful and versatile SE model. To this end, we propose LLaSE-G1, a LLaMA-based language model that incentivizes generalization capabilities across various SE tasks. The architecture of LLaSE-G1 is simple yet effective, consisting of a WavLM~\cite{wavlm} encoder for feature extraction, a LLaMA-based language model for token prediction, and an X-codec2~\cite{llasa} decoder for waveform reconstruction. Specifically, to address the acoustic inconsistency caused by the information loss inherent in discrete tokens, we replace the discrete token inputs with continuous representations extracted from the WavLM encoder and predict speech tokens obtained from X-codec2. The WavLM encoder provides sufficient speech details, and X-codec2 integrates semantic and acoustic features into speech tokens, thus maximizing acoustic preservation. Additionally, to incentivize generalization, LLaSE-G1 utilizes dual-channel inputs and outputs, unifying the degraded speech and optional reference signals and constraining all tasks under a cross-entropy loss function. Through extensive experiments, LLaSE-G1 demonstrates superior performance on NS, PLC, TSE, and AEC benchmarks. Furthermore, LLaSE-G1 exhibits emergent capabilities for previously unseen SE tasks, such as SS, and shows scaling effects at test time, where performance improves with increased compute.

% In this paper, we argue that with appropriate design, a single language model can be a powerful and versatile SE model. To this end, we propose LLaSE-G1, a LLaMA-based LM that incentivizes generalization capability for various SE tasks. LLaSE-G1 keeps a simple structure, which consists of a WavLM encoder for feature extraction, a LLaMA-based LM for token prediction, and a X-codec2~\cite{llasa} decoder for waveform reconstruction. Specifically, to address acoustic inconsistency caused by information loss of discrete tokens, we replace discrete token inputs with continuous representations extracted from the WavLM encoder and predict speech tokens obtained from X-codec2, maximizing acoustic preservation. Besides, to incentivize generalization capability, LLaSE-G1 dedicates dual-channel inputs and outputs, which unify inputs of degraded speech and reference signal and constrain all tasks under the cross-entropy loss. Through extensive experiments, LLaSE-G1 demonstrates superior performance on NS, PLC, TSE, and AEC benchmarks. Moreover, LLaSE-G1 exhibits emergent capabilities for unseen SE tasks, such as SS, and scaling effects during test time where the performance of some tasks is improved when scaling test-time compute.

In summary, our paper makes several key contributions:

\begin{itemize}
    \item We propose LLaSE-G1, a LLaMA-based language model that incentivizes generalization capability for speech enhancement.
    \item We effectively address the acoustic inconsistency by leveraging both continuous and discrete representations, and we design dual-channel inputs and outputs, which unify various SE tasks without the need for task IDs. Notably, AEC, PLC, and SS tasks are being introduced to generative models for the first time.
    \item LLaSE-G1 outperforms existing models on several SE benchmarks and demonstrates scaling effects during test time and emergent capabilities for unseen SE tasks. We release the codes and checkpoints as open-source.
    
\end{itemize}

\section{Related Work}
Speech enhancement refers to the technology of recovering high-quality target speech from degraded speech, which includes multiple subtasks~\cite{DBLP:journals/taslp/WangC18a,VoiceFixer}. Traditional speech enhancement, which relies on statistical analysis and signal processing, often struggles with generalization in dynamic scenarios. With the development of deep learning, data-driven speech enhancement has become the mainstream approach and can be divided into two categories: discriminative SE and generative SE~\cite{DBLP:conf/icassp/LemercierRWG23}. Discriminative SE models learn a mapping between degraded speech and the corresponding clean speech targets, including methods such as time-frequency (T-F) masking~\cite{DBLP:journals/taslp/WilliamsonW17} and time-domain approaches~\cite{DBLP:conf/interspeech/0004M18}. In contrast, generative models reconstruct the clean speech by learning the data distribution of the target speech, such as diffusion-based generative models~\cite{DBLP:journals/corr/abs-2303-13336,DBLP:journals/taslp/RichterWLLG23}. Recently, researchers have also begun to explore the use of LMs to improve generative speech enhancement~\cite{wang2024selmspeechenhancementusing, yao2025gensegenerativespeechenhancement}.

% Speech enhancement models can generally be divided into two categories, discriminative speech enhancement models and generative speech enhancement models. discriminative SE models achieve enhancement by mapping the spectrum, while generative SE models attain improvement by reconstructing target information. Both types of models have demonstrated substantial capabilities in addressing single-task problems.

\subsection{Discriminative Speech Enhancement}
Traditionally, speech enhancement encompasses tasks such as NS, PLC, TSE, AEC, and SS, with NS also requiring dereverberation.
For different tasks, discriminative models often have different architectures. In NS tasks, most models are based on the convolutional encoder-decoder (CED) architecture. 
% DCCRN~\cite{hu2020dccrndeepcomplexconvolution} introduced a complex convolutional recurrent network (CRN), utilizing complex-valued convolution and LSTM to process complex spectra. 
FRCRN~\cite{zhao2024frcrnboostingfeaturerepresentation} adds a frequency recurrent network to the CED architecture, achieving excellent performance. 
In PLC tasks, Generative Adversarial Networks (GANs) are commonly used. 
% Li et al.~\cite{DBLP:conf/interspeech/LiZZGY22} employ a time domain generator and combine time domain and time-frequency domain discriminators for adversarial training. 
BS-PLCNet~\cite{zhang2024bsplcnetbandsplitpacketloss} uses multitask learning and multi discriminators, winning the latest PLC Challenge~\cite{2024plc}.
In TSE tasks, the speaker embedding paradigm is widely adopted. TSE approaches usually use speaker verification models~\cite{DBLP:conf/interspeech/DesplanquesTD20,DBLP:conf/icassp/WangLWCZXDQ23} to extract embeddings from enrollment audio and integrate into noise suppression networks. This approach has been successful in the personalized tracks of the Deep Noise Suppression challenges, as demonstrated by the TEA-PSE series models~\cite{ju2023teapse30tencentetherealaudiolabpersonalized,DBLP:conf/icassp/JuRYFLCWXS22}. 
For AEC tasks, an important issue is how to deal with the delay estimation and alignment between reference signals and microphone signals. 
% Hybrid approaches use signal processing methods for delay estimation and linear filtering, with neural network models serving as post-filter for residual echo cancellation, such as GFTNN~\cite{DBLP:conf/icassp/ZhangWSFTFX22} and MTFAANet~\cite{DBLP:conf/icassp/ZhangWYW22}. 
DeepVQE~\cite{indenbom2023deepvqerealtimedeep} utilizes attention-based delay estimation, employing fully neural networks to solve echo cancellation problems.  
For SS tasks, common models such as TF-GridNet~\cite{DBLP:journals/taslp/WangCCLKW23} and Mossformer2~\cite{DBLP:conf/icassp/ZhaoMNZ000YN024} can only handle a fixed number of speakers. SepTDA~\cite{DBLP:conf/icassp/LeeCKW024} introduces a transformer decoder-based attractor, capable of handling a dynamic number of speakers, but still requires specifying the maximum number of speakers. 

While these discriminative models have achieved excellent performance across various tasks, their generalization ability is limited by the availability of training data and model parameters~\cite{DBLP:conf/interspeech/WelkerRG22}. This can lead to performance degradation in unseen scenarios. Additionally, these models may introduce undesired speech distortion and phonetic inaccuracies to enhanced speech~\cite{DBLP:journals/taslp/WangTW20}.

\subsection{Generative Speech Enhancement}
% MaskSR~\cite{li2024masksrmaskedlanguagemodel} uses a mask generation model to simultaneously handle noise, reverberation, clipping, and bandwidth limitation. AnyEnhance~\cite{zhang2025AnyEnhanceunifiedgenerativemodel} introduces a prompt-guidance mechanism, enabling in-context learning capabilities to further perform target speaker extraction tasks. However, echo cancellation, which requires reference audio input, has not been included in the multitasking framework.

% In recent years, generative SE models such as GANs, VAEs, and diffusion models have been applied to speech enhancement tasks~\cite{pascual2017seganspeechenhancementgenerative,Fang_2021}.
Early generative SE primarily relied on GANs and VAEs~\cite{pascual2017seganspeechenhancementgenerative,Fang_2021}. Although these approaches offered new perspectives, they still did not surpass the performance of discriminative models. In recent years, diffusion-based generative models have been applied to speech enhancement. CDiffusion~\cite{DBLP:conf/icassp/LuWWRYT22} defines the conditional diffusion process by incorporating noisy data into the diffusion process. DiffSep~\cite{DBLP:conf/icassp/ScheiblerJCBCC23} designs stochastic differential equations (SDE)~\cite{DBLP:conf/iclr/0011SKKEP21}. By solving the corresponding reverse-time SDE, it is possible to recover individual sources from the mixture. Despite diffusion models achieving superior speech quality over discriminative models in noise suppression (NS) and source separation (SS), these tasks were previously independent of each other, requiring separate training of different models, and proving difficult to generalize to other tasks.

Recently, researchers have begun to explore the use of a joint framework, leveraging the capabilities of generative models to integrate multiple enhancement tasks into a single model. Nemo~\cite{DBLP:journals/corr/abs-2409-16117} and SpeechFlow~\cite{DBLP:conf/iclr/Liu0VSTH24} pre-trained on large-scale datasets and can be adapted to downstream tasks such as NS and TSE through fine-tuning. AnyEnhance~\cite{zhang2025AnyEnhanceunifiedgenerativemodel} achieves both NS and TSE without the need for fine-tuning. It introduces a prompt guidance mechanism, enabling in-context learning capabilities.

With the rise of LMs in their ability to handle multiple tasks, LMs have also been introduced into speech enhancement. SELM~\cite{wang2024selmspeechenhancementusing} employs a WavLM-based k-means tokenizer and predicts clean tokens from noisy tokens, marking the first introduction of LMs into the NS domain. 
% TSELM~\cite{DBLP:journals/corr/abs-2409-07841}, by incorporating speaker embeddings, has demonstrated the feasibility of task generalization.
MaskSR~\cite{li2024masksrmaskedlanguagemodel} uses a mask generation model to simultaneously handle noise, reverberation, clipping, and bandwidth limitation. However, existing unified enhancement models have not considered the echo cancellation task, which requires reference audio input and the need to address delay estimation and alignment issues. We suggest that by leveraging the powerful modeling capabilities of LMs, it is possible to develop a general speech enhancement model that unifies NS, PLC, TSE, AEC and SS.

\section{LLaSE-G1}

\subsection{Overall Architecture}
\begin{figure}[ht]
  \centering
  \includegraphics[width=\linewidth]{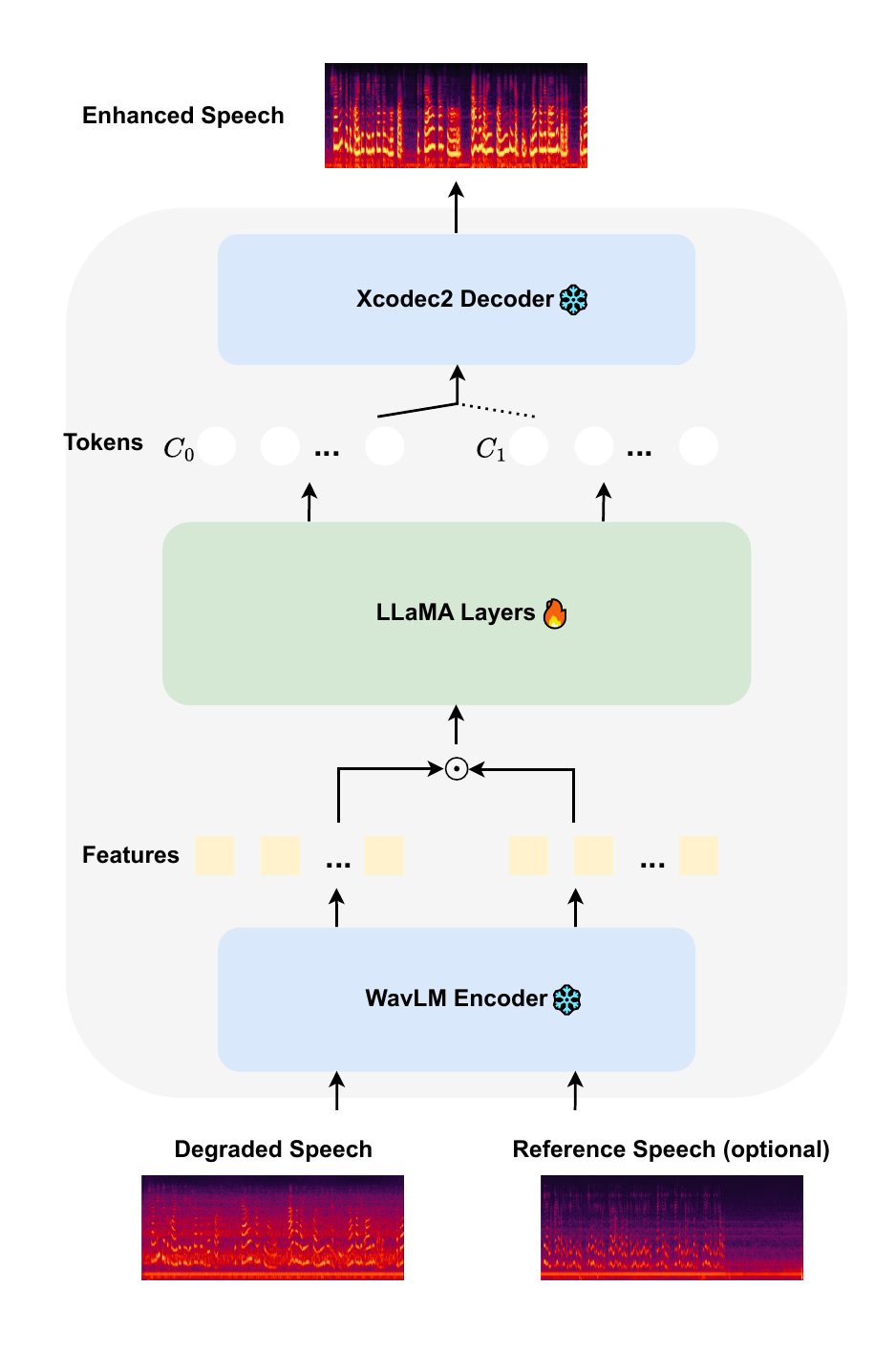}
  \caption {Overall Architecture of LLaSE-G1. LLaSE-G1 simplifies model architecture to support various SE tasks.}
  \label{fig:Architecture}
\end{figure}

LLaSE-G1 is designed to incentivize generalization across various SE tasks with a single LLaMA-based LM~\cite{grattafiori2024llama3herdmodels}. As shown in Figure~\ref{fig:Architecture}, compared to previous specialist models such as FRCRN~\cite{zhao2024frcrnboostingfeaturerepresentation}, TEA-PSE 3.0~\cite{ju2023teapse30tencentetherealaudiolabpersonalized}, Align-ULCNet~\cite{shetu2024alignulcnetlowcomplexityrobustacoustic}, mossformer2~\cite{zhao2024mossformer2combiningtransformerrnnfree} and BS-PLCNet~\cite{zhang2024bsplcnetbandsplitpacketloss}, LLaSE-G1 greatly simplifies the model structure, keeping three main components: (1) a WavLM encoder, (2) a LLaMA-based LM and (3) an X-codec2 decoder. Specifically, the WavLM encoder extracts continuous speech features from degraded speech. The LLaMA-based LM takes speech features as input and predicts discrete speech tokens extracted by X-codec2 in an autoregressive manner. Finally, the X-codec2 decoder reconstructs enhanced speech from predicted speech tokens.

% Formally, let:
% 1. \(\texttt{Vertorize}(X) = \{x_1, \dots, x_N\} \) be the WavLM encoder, which converts input degraded speech \(X\) into \(N\) speech features.
% 2. \( \texttt{Vertorize}(P) = \{p_1, \dots, p_T\} \) be the WavLM encoder, which converts optional reference speech \(X\) into \(T\) speech features.
% 3. \( \texttt{Tokenize}(Y) = \{y_1, \dots, y_M\} \) be the X-codec2 encoder, which converts an enhanced speech \(Y\) into \(M\) speech tokens.
% 4. \( \texttt{Detokenize}(\{y_1, \dots, y_M\}) = \hat{Y} \) be the X-codec2 decoder, which reconstructs the waveform \(\hat{Y}\) from its token representations. As the downsampling rate of WavLM is the same as that of X-codec2, $N$ is equal to $M$.
% Given a dataset \(\mathcal{D} = \{(X_i, P_i, Y_i)\}_{i=1}^S\), where \(X_i\) is the degraded speech, \(Y_i\) is the enhanced speech and \(P_i\) is the reference speech or empty if unavailable, we represent each pair \( (X_i, P_i, Y_i)\) as a sequence \( (x_1, \dots, x_N, p_1, \dots, p_T, y_1, \dots, y_M)\). 
% Since the \(X_i\) and \(P_i\) are always given as input during training and inference, we pad \(X_i\) and \(P_i\) to the same length and the LM $\theta$ focuses on learning the conditional probability:
% \begin{equation}
%     \begin{aligned}
%         &P (x_1, \ldots, x_N, p_1, \ldots, p_T, y_1, \ldots, y_M; \theta) \\
%     =& \prod_{j=1}^M P (y_j | x_1 \odot p_1, \ldots, x_j\odot p_j; \theta),
%     \end{aligned}
% \end{equation}
% where $\odot$ is the concatenation between $x$ and $p$ in the channel axis.

Formally, let:
1. \( \texttt{Vertorize}(X) = \{x_1, \dots, x_N\} \) be the WavLM encoder, which converts input degraded speech \(X\) into \(N\) speech features.
2. \( \texttt{Vertorize}(P) = \{p_1, \dots, p_T\} \) be the WavLM encoder, which converts optional reference speech \(P\) into \(T\) speech features.
3. \( \texttt{Tokenize}(Y) = \{y_1, \dots, y_M\} \) be the X-codec2 encoder, which converts an enhanced speech \(Y\) into \(M\) speech tokens.
4. \( \texttt{Detokenize}(\{y_1, \dots, y_M\}) = \hat{Y} \) be the X-codec2 decoder, which reconstructs the waveform \(\hat{Y}\) from its token representations. As the downsampling rate of WavLM is the same as that of X-codec2, $N$ is equal to $M$.
Given a dataset \(\mathcal{D} = \{(X_i, P_i, Y_i)\}_{i=1}^S\), where \(X_i\) is the degraded speech, \(Y_i\) is the enhanced speech and \(P_i\) is the reference speech or empty if unavailable, we represent each pair \((X_i, P_i, Y_i)\) as a sequence \((x_1, \dots, x_N, p_1, \dots, p_T, y_1, \dots, y_M)\). 
Since the \(X_i\) and \(P_i\) are always given as input during training and inference, we pad \(X_i\) and \(P_i\) to the same length and the LM $\theta$ focuses on learning the conditional probability:
\begin{equation}
    \begin{aligned}
        &P(x_1, \ldots, x_N, p_1, \ldots, p_T, y_1, \ldots, y_M; \theta) \\
    =& \prod_{j=1}^M P(y_j | x_1 \odot p_1, \ldots, x_j\odot p_j; \theta),
    \end{aligned}
\end{equation}
where $\odot$ is the concatenation between $x$ and $p$ in the channel axis.

% Our system takes only two essential inputs and does not include any extra prompts. The first input is the audio to be processed, while the optional second input can be either the enrollment audio of the target speaker for the TSE task or reference audio for the AEC task.

% The two input streams will first be transformed into continuous embedding input features by WavLM and then concatenated along the feature dimension before being sent to the LM. The LM used in our system is based on LLaMA architecture, and we set its prediction target to be the clean label tokens produced by the X-codec2 encoder. Lastly, the X-codec2 Decoder is used to reconstruct the audio signal from the tokens predicted by LM.

\subsection{Maximizing Acoustic Preservation}

As highlighted by WavChat~\cite{ji2024wavchatsurveyspokendialogue}, speech representations can be broadly categorized into two types: continuous and discrete representations. Continuous representations, typically extracted from self-supervised learning (SSL) models like HuBERT~\cite{hsu2021hubertselfsupervisedspeechrepresentation} and WavLM~\cite{wavlm}, are considered lossless carriers for speech, capturing intricate speech details. In contrast, discrete representations are derived from speech codecs such as Encodec~\cite{encodec} and DAC~\cite{dac}, which, while facilitating language modeling, are lossy due to the information loss during quantization.
To address this, LLaSE-G1 adopts continuous representations as input and predicts discrete representations, aiming to maximize acoustic preservation throughout the enhancement process.

For continuous speech representations, we utilize WavLM as the extractor. WavLM is an SSL model that combines a convolutional feature encoder with a transformer encoder. Pre-trained on large-scale speech data, it excels across various speech-processing tasks. Previous research~\cite{knnvc, vectok} has shown that features extracted from the 6th layer of WavLM contain sufficient acoustic information for high-fidelity speech reconstruction. Therefore, we leverage the features from this layer as the input representations for the language model.

For discrete speech representations, we use X-Codec2 as the extractor. X-Codec2 is a recently developed speech codec that integrates semantic and acoustic features into a unified codebook, ensuring a 1D causal dependency. This design reflects the inherent left-to-right temporal structure of audio signals, while also preserving more acoustic information compared to traditional 1D semantic tokens. Consequently, we adopt the speech tokens extracted by X-Codec2 as the modeling target for the language model.

\subsection{Unifying Various SE tasks}
Although different speech enhancement tasks are applied in various scenarios, they share underlying commonalities, such as the need to determine which components should be removed from the noisy speech. To this end, LLaSE-G1 employs a dual-channel input and output framework that unifies several SE tasks within a single language model (LM). These tasks include NS, TSE, PLC, AEC, and SS.

Systematically, NS, PLC, and SS require only the degraded speech as input, while TSE and AEC need both the degraded speech and an additional reference speech. In contrast to previous PLC models, we do not use the lossy labels that indicate missing speech frames, thereby simplifying the data requirements. 

To unify the input representations, we introduce a dual-channel input: one channel for the degraded speech and the other for the optional reference speech. These representations are padded to the same length and concatenated along the channel dimension. Notably, if the reference speech is unavailable, we set the second channel to zero.

While NS, PLC, AEC, and TSE tasks output a single enhanced speech, we note that AEC requires the removal of information related to the reference speech, while TSE necessitates the preservation of reference speech information. To address this, we introduce a dual-channel output with two linear projection heads to unify the output representations. The output embedding of the LLM is passed through these two linear projection to produce two embeddings from it. The first channel $c_0$ predicts tokens related to reference speech and the second channel $c_1$ predicts tokens irrelevant to reference speech. With these designs, for tasks including NS, AEC, and PLC, we employ a single-supervision strategy $\mathcal{L}_{\text{S}}$ through the cross-entropy loss between $c_0$ and the tokens $t_0$ extracted from the clean signal:
\begin{equation}
\mathcal{L}_{\text{S}} = -\frac{1}{N} \sum_{k=1}^N t_0^{ (k)} \log \left (c_0^{ (k)}\right)
\end{equation}
For the TSE task, we implement a dual-supervision strategy $\mathcal{L}_{\text{D}}$ with separate constraints for both outputs. The first output $c_0$ handles interfering speaker, while the second output $c_1$ is dedicated to target speaker extraction. The $\mathcal{L}_{\text{D}}$ is formulated as:
\begin{equation}
\mathcal{L}_{\text{D}} =  -\frac{1}{N} \sum_{k=1}^N [ t_0^{ (k)} \log \left (c_0^{ (k)}\right) + t_1^{ (k)} \log \left (c_1^{ (k)}\right)]
\end{equation}
Importantly, in LLaSE-G1, SS is treated as an unseen task throughout the entire training process.

% As for TSE and AEC, the role of reference audio is actually the opposite. Therefore, placing both the output and the constraints on the same track could make it difficult for the model to converge. As a result, the target output for TSE is $s_0$ separately, while the target output for AEC and other tasks is $s_1$.

% We use X-codec2's encoder to generate target tokens from clean speech labels for LM.
% For tasks including NS, AEC, PLC, we employ a single-target cross-entropy loss based on the predicted tokens for the clean signal:
% \[
% \mathcal{L}_{\text{s1}} = -\frac{1}{N} \sum_{k=1}^N t_1^{ (k)} \log \left (s_1^{ (k)}\right)
% \]

\section{Experiments and Results}
\subsection{Experimental Setup}
% Here we briefly introduce the experimental setup. Details of the setup can be found in the Appendix~\label{sec:appendixA}.

\textbf{Datasets.} For the training data, we use the Librispeech, HiFiTTS, and DNS Challenge datasets~\cite{2020dns, 2023dns}, along with internal datasets as original clean speech, totalling approximately 5000 hours. Room impulse responses (RIRs) are sourced from the DNS Challenge datasets. The noise data contains nearly 1000 hours, sourced from DEMAND, ESC-50, DNS Challenge, AEC Challenge~\cite{2023aec}, and internal datasets. 
% Details of the data simulation configurations and test set for each subtask are provided in Appendix~\ref{sec:appendix1}.

\textbf{Data augmentation.} We utilized dynamic data augmentation during training.
For the NS task, the clean audio and noise are randomly selected and mixed with a signal-to-noise Ratio (SNR) ranging from [-5,20] dB. Both clean and noisy signals have a 50\% probability of adding reverberation. 
% The noisy signal is then scaled by a factor between 0.3 and 0.9, ensuring its amplitude matches that of the clean signal.
In the PLC task, we use a two-state first-order Markov chain to describe the packet loss status of the current frame and the next frame. The transition and hold probabilities for Markov states are selected between 0.05 and 0.95. 
% The packet loss rate is set to not exceed 50\%. 
We directly generate a binary mask sequence and apply it to the clean speech.
For the AEC task, we randomly select a real echo signal and its corresponding reference signal from the far-end single talk in the AEC Challenge dataset. The signal-to-echo ratio (SER) ranges from -15 dB to 15 dB. Noise is added with a 20\% chance, and the SNR is between -5 dB and 20 dB. 
% The probability of encountering both far-end only and near-end only signals is set at 2.5\%.
For the Target Speaker Extraction (TSE) task, we select a clean speech segment and its corresponding auxiliary segment for the enrollment speech, while a different speaker is chosen for the interference speech. There is a 5\% probability that no interfering speaker is present. Target speech and interfering speech are mixed with an SNR ranging from [-15, 15] dB, with an additional 10\% probability of adding extra noise.

The audio length for each batch is 8 seconds. Before being fed into the model, the audio is randomly truncated to a length between 4 and 8 seconds to ensure the model's ability to generalize to different audio lengths. During training, the distribution of tasks (NS, PLC, AEC, and TSE) is evenly balanced. Within each batch, the data are of the same task type. Gradient accumulation is enabled to help the model adapt to multi-task learning, with parameter updates occurring every 20 steps.

\textbf{Model configuration.} We use the open-source checkpoints of WavLM-large\footnote{\href{https://huggingface.co/microsoft/wavlm-large}{WavLM-Large on Hugging Face}} and X-codec2\footnote{\href{https://huggingface.co/HKUSTAudio/xcodec2}{X-codec2 on Hugging Face}}. The LLaMA-based LM comprises 16 LLaMA layers, each with 16 attention heads, a dropout rate of 0.1, a hidden size of 2048, and an intermediate size of 4096. The total number of parameters in the model is approximately 1.07 billion. More details are given in Appendix~\ref{sec:appendix2}. 
% We trained the model for 100,000 steps using 4 NVIDIA L40 GPUs, with a batch size of 6 per GPU and the AdamW optimizer. The learning rate is set to 1e-4. More detailed model configurations are provided in Appendix~\ref{sec:appendix2}.

\textbf{Baseline systems.} We evaluate the performance of our LLaSE-G1 with several state-of-the-art (SOTA) models of each subtask, including the winners of the recent signal processing grand challenges~\cite{2020dns, 2023dns,2023aec,2024plc,2022plc} for each task. Details of the baseline system and test set for each subtask are provided in Appendix~\ref{sec:appendix1}.

\textbf{Evaluation Metrics.} We use objective metrics to evaluate the performance of the baseline systems and our model. DNSMOS~\cite{DNSMOS} include speech quality (SIG), background noise quality (BAK), and overall quality (OVRL) of the audio. AECMOS~\cite{AECMOS} consists of echo annoyance MOS (EMOS) and other degradation MOS (DMOS). PLCMOS~\cite{PLCMOS} is used to assess the quality of audio processed by PLC algorithms. All MOS scores range from 1 to 5, representing audio quality from low to high. 
SpeechBERTScore (SBS)~\cite{saeki2024speechbertscorereferenceawareautomaticevaluation} is also employed to evaluate the semantic similarity between the enhanced audio and the reference audio. Following~\cite{zhang2025AnyEnhanceunifiedgenerativemodel}, we use HuBERT-base\footnote{\href{https://huggingface.co/facebook/hubert-base-ls960}{Hubert-base on Hugging Face}} model to extract semantic features. 
For acoustic similarity, we calculate speaker similarity SimW$_B$ based on the WavLM-base-sv model\footnote{\href{https://huggingface.co/microsoft/wavlm-base-plus-sv}{WavLM-base-sv on Huggingface}} to evaluate the performance. Subjective evaluations are also conducted using the Mean Opinion Score (MOS) as the primary metric to assess the model's performance.

% finetuned WavLM-TDCNN speaker verification model\footnote{\href{https://github.com/microsoft/UniSpeech/tree/main/downstreams/speaker_verification}{WavLM-TDCNN on Github}} and
% For the NS task, we use DNSMOS; for the TSE task, we use pDNSMOS; for the SS task, we use DNSMOS and Word Error Rate (WER). For the AEC task, we use AECMOS, and for the PLC task, we use DNSMOS and PLCMOS.

\textbf{Inference.} 
For each task, we conduct single and multiple inferences. For multiple inferences, we infer 10 times and take the best result where the model’s output is used as the input for the next inference.
For the PLC task, we only employ the audio to be processed as input, without the lossy label. For the TSE task, we keep the enrollment audio unchanged during multiple inferences. For the AEC task, we only use the reference audio for the first inference, subsequent inferences are treated as NS tasks. 
For the SS task, we employ a two-stage inference strategy. First, we separate one speaker from the mixed audio, and then use the first separated speaker's audio as a reference for the second inference stage.
% The SS task was not included in the training data and represents an unseen task. 
% We use a single NVIDIA L40 GPU for inference and take the best results of multiple references as our results.

% We use NVIDIA L40 GPUs for both training and inference, taking the best results from multiple references as our final output. Specific details regarding training, inference, and model configuration are provided in Appendix A.
    
\subsection{Experimental Results}
\subsubsection{Noise Suppression}
Table~\ref{tab:Noise Suppression} presents a comparison between the proposed LLaSE-G1 and several state-of-the-art (SOTA) discriminative and generative models. The "With Reverb" column corresponds to the test set containing reverberation, while the "No Reverb" column refers to the clean test set without reverberation. The results show that generative noise suppression (NS) models consistently outperform discriminative ones, particularly under reverberant conditions. Even with single inference, LLaSE-G1 surpasses most competing systems. When employing multiple inferences, its performance improves further, achieving a SOTA OVRL score of 3.49 on the "No Reverb" test set and 3.42 on the "With reverb" test set. 

Notably, LLaSE-G1 demonstrates exceptional performance on the with reverb test set compared to other generative enhancement systems, further highlighting the efficiency of continuous representations and the Xcodec2 decoder in handling challenging noise suppression tasks.

\begin{table}[ht]
    \centering
    \caption{DNSMOS scores on the Interspeech 2020 DNS Challenge blind test set. "D" represents Discriminative and "G" represents Generative. LLaSE-G1\(_{\text{single}}\) and LLaSE-G1\(_{\text{multi}}\) represent single inference and multiple inference using LLaSE-G1, respectively.}
    \footnotesize
    \setlength\tabcolsep{1pt}
    \begin{tabular}{ll ccc ccc}
        \toprule
        Model & Type & \multicolumn{3}{c}{With Reverb} & \multicolumn{3}{c}{No Reverb} \\
        \cmidrule (lr){3-5} \cmidrule (lr){6-8}
        & & SIG & BAK & OVRL & SIG & BAK & OVRL \\
        \midrule
        Noisy    & -  & 1.76 & 1.50 & 1.39 & 3.39 & 2.62 & 2.48 \\
        \hline
        Conv-TasNet %~\cite{Luo_2019}    
        & D  & 2.42 & 2.71 & 2.01 & 3.09 & 3.34 & 3.00 \\
        DEMUCS %~\cite{défossez2019demucsdeepextractormusic}        
        & D  & 2.86 & 3.90 & 2.55 & 3.58 & 4.15 & 3.35 \\
        FRCRN %~\cite{zhao2024frcrnboostingfeaturerepresentation}          
        & D  & 2.93 & 2.92 & 2.28 & 3.58 & 4.13 & 3.34 \\
        \hline
        SELM %~\cite{wang2024selmspeechenhancementusing}           
        & G  & 3.16 & 3.58 & 2.70 & 3.51 & 4.10 & 3.26 \\
        MaskSR %~\cite{li2024masksrmaskedlanguagemodel}         
        & G  & 3.53 & 4.07 & 3.25 & 3.59 & 4.12 & 3.34 \\
        AnyEnhance %~\cite{zhang2025AnyEnhanceunifiedgenerativemodel}       
        & G  & 3.50 & 4.04 & 3.20 & 3.64 & 4.18 & 3.42 \\
        GenSE %~\cite{yao2025gensegenerativespeechenhancement}          
        & G  & 3.49 & 3.73 & 3.19 & 3.65 & 4.18 & 3.43 \\
        \hline
        LLaSE-G1\textsubscript{single} & G  & 3.59 & 4.10 & 3.33 & 3.66 & 4.17 & 3.42 \\
        LLaSE-G1\textsubscript{multi} & G  & \textbf{3.65} & \textbf{4.16} & \textbf{3.42} & \textbf{3.71} & \textbf{4.19} & \textbf{3.49} \\

        \bottomrule
    \end{tabular}
    \label{tab:Noise Suppression}
\end{table}

% Our model already surpasses other baseline systems with a single inference. After multiple inferences, its performance improves further, achieving a new state-of-the-art result.

% However, discriminative SE models still have advantages in semantic similarity. Among generative SE models, both AnyEnhance and our model achieve relatively high semantic similarity scores on the no\_reverb test set, even outperforming some discriminative SE models. Furthermore, on the with\_reverb test set, our model outperforms other generative models and achieves results at 0.77 that are very close to those of discriminative models at 0.78, giving strong evidence to the acoustic preservation capability of our model.

\subsubsection{Packet Loss Concealment}

We compared LLaSE-G1 with the top-performing models~\cite{zhang2024bsplcnetbandsplitpacketloss,2022kuaishou,liu2022plcnet,LPCNet} from the most recent two challenges on the Interspeech 2022 PLC blind test set~\cite{2022plc}. 
%For comparison, we use BS-PLCNet ~\cite{zhang2024bsplcnetbandsplitpacketloss}, Team Kuaishow~\cite{2022kuaishou}, which are the winners of the  2024 challenge and 2022 challenge respectively, and other systems in challenge like PLCNet~\cite{liu2022plcnet} and LPCNet~\cite{LPCNet}  as our baseline systems. 
% Compared to the baseline systems ~\cite{zhang2024bsplcnetbandsplitpacketloss,2022kuaishou,liu2022plcnet,LPCNet}, LLaSE-G1 automatically identifies the loss mask, while the baseline system requires an additional lossy mask.
It is important to note that LLaSE-G1 operates as a blind PLC without the need for lossy labels. This means LLaSE-G1 autonomously determines whether to perform PLC without prior knowledge of which frames experienced packet loss, making it a more challenging task and distinct from the models participating in the PLC Challenge.

\begin{table}[ht]
    \centering
    \caption{DNSMOS OVRL and PLCMOS scores on ICASSP 2022 PLC-challenge blind testet.}
    \footnotesize
    \setlength\tabcolsep{3pt}
    \begin{tabular}{ll cc}
        \toprule
        Model & Type  & {OVRL} & {PLCMOS} \\
        \midrule
        Noisy    & -  & 2.56 & 2.90 \\
        \hline
        KuaishouNet   & D  &  -  & 4.27\\
        LPCNet   & D  &   3.09   & 3.74\\
        PLCNet   & D  &   -    & 3.83\\
        BS-PLCNet  & D  & 3.20 & 4.29 \\
        \hline
        LLaSE-G1\textsubscript{single} & G  & 3.03 & 3.68 \\
        LLaSE-G1\textsubscript{multi} & G  & \textbf{3.27} & \textbf{4.30} \\
        \bottomrule
    \end{tabular}
    \label{tab:PLC scores}
\end{table}

The results in Table~\ref{tab:PLC scores} demonstrate significant improvement with our model through inference time scaling. Specifically, the multi-inference approach boosts both OVRL and PLCMOS scores, with OVRL increasing from 3.03 to 3.27 and PLCMOS rising from 3.68 to 4.30, highlighting its effectiveness. LLaSE-G1's results on blind PLC surpassed those of other models using informed PLC, demonstrating the powerful content understanding and generation capabilities of LMs.
% Furthermore, LLaSE-G1 achieves these gains without relying on the extra lossy mask used by the baselines, emphasizing the model's efficiency and generalization.

\subsubsection{Target Speaker Extraction}

We use the ICASSP 2023 DNS blind test set~\cite{2023dns} for the TSE task evaluation, which includes two tracks: the headset track and the speakerphone track. 
% We employ pDNSMOS~\cite{2023dns} as our evaluation metric. % We compare our model with two baseline systems: TEA-PSE 3.0~\cite{ju2023teapse30tencentetherealaudiolabpersonalized}, the winner of the challenge, and NAPSE~\cite{yan2023npuelevocpersonalizedspeechenhancement}, which placed second.

\begin{table}[ht]
    \centering
    \caption{pDNSMOS scores on ICASSP 2023 DNS-challenge
blind testet. }
    \footnotesize
    \setlength\tabcolsep{1.5pt}
    \begin{tabular}{ll ccc ccc}
        \toprule
        Model & Type & \multicolumn{3}{c}{Track 1} & \multicolumn{3}{c}{Track 2} \\
        \cmidrule (lr){3-5} \cmidrule (lr){6-8}
        & & SIG & BAK & OVRL & SIG & BAK & OVRL \\
        \midrule
        Noisy    & -  &4.15 & 2.37 & 2.71 & 4.05 & 2.16 & 2.50 \\
        \hline
        TEA-PSE 3.0 %~\cite{ju2023teapse30tencentetherealaudiolabpersonalized}  
        & D  &  4.12 & \textbf{4.05} & 3.65 & 3.99 & \textbf{3.95} & 3.49 \\
        NAPSE %~\cite{yan2023npuelevocpersonalizedspeechenhancement} 
        & D  &  3.81 &  3.99  &  3.38 & 3.92 & 4.17 & 3.56 \\
        \hline
        LLaSE-G1\textsubscript{single} & G  & \textbf{4.21} & 3.99 & \textbf{3.72} & 4.08 & 3.84 & 3.55\\
        LLaSE-G1\textsubscript{multi} & G  & 4.20 & 3.97 & 3.70 & \textbf{4.11} & 3.86 & \textbf{3.58} \\
        
        \bottomrule
    \end{tabular}
    \label{tab:TSE pdnsmos}
\end{table}

As shown in Table~\ref{tab:TSE pdnsmos}, LLaSE-G1 consistently achieves significantly higher SIG MOS scores across both tracks, surpassing all other methods. This indicates that language model-based generative approaches offer higher audio quality with reduced signal distortion. While TEA-PSE 3.0 and NAPSE exhibit certain advantages on headset and speakerphone devices respectively, LLaSE-G1 delivers the best overall performance across both tracks, demonstrating superior device generalization compared to discriminative models.

% Meanwhile, multiple inferences did not significantly impact the TSE task, as a single inference already achieved OVRL scores of 3.72 and 3.55.
% LLaSE-G1 outperforms the challenge winner and other baseline systems in both Track 1 and Track 2. In Track 1, LLaSE-G1 achieves the highest OVRL score of 3.72 in the single inference setting and 3.70 in multi-inference. In Track 2, LLaSE-G1 performs similarly well, with an OVRL score of 3.55 in the single inference and 3.58 in multi-inference, outperforming all other models. These results demonstrate the effectiveness of LLaSE-G1 in enhancing both signal and background noise in more complex scenarios.

\subsubsection{Acoustic Echo Cancellation}
LLaSE-G1 is the first generative model to integrate the AEC task into a unified framework.
% We use the ICASSP 2023 AEC-challenge blind testet~\cite{2023aec} for testing. % For baseline comparison, we choose recent efficient and state-of-the-art systems as our baseline systems.~\cite{shetu2024ulcnet,shetu2024alignulcnetlowcomplexityrobustacoustic,indenbom2023AlignCruse}. And DeepVQE~\cite{indenbom2023deepvqerealtimedeep}is the state-of-the-art model in the AEC task.
As shown in Table~\ref{tab:AEC}, LLaSE-G1 demonstrates comparable performance to the SOTA discriminative AEC approaches, showcasing the potential of LMs-based generative models for the AEC task.
% our model outperforms several discriminative SE models and achieves results comparable to the state-of-the-art system. 
% Our results demonstrate the potential of generative SE models for the acoustic echo cancellation task.
\begin{table}[ht]
    \centering
    \caption{AECMOS Echo (EMOS) and Degradation (DMOS) scores on ICASSP 2023 AEC-challenge blind test set."DT" represents double-talk, FEST means far-end only and NEST means near-end only.}
    \footnotesize
    \setlength\tabcolsep{2pt}
    \begin{tabular}{ll cc c c}
        \toprule
        Model & Type & \multicolumn{2}{c}{DT} & \multicolumn{1}{c}{FEST} & \multicolumn{1}{c}{NEST} \\
        & & EMOS & DMOS & EMOS & DMOS \\
        \midrule
        Align-CRUSE & D  & 4.60 & 3.95 &  4.56 & - \\
        DeepVQE & D & \textbf{4.70} & \textbf{4.29} & 4.69 & \textbf{4.41} \\
        ULCNetAENR  & D  &  4.54 & 3.58 & 4.73 & 4.15  \\
        Align-ULCNet  & D  &   4.60 &  3.80 & \textbf{4.77} & 4.28  \\
        \hline
        LLaSE-G1\textsubscript{single}  & G  & 4.42 & 3.82 & 4.64 & 3.66 \\
        LLaSE-G1\textsubscript{multi} & G  & 4.52 & 3.91 & 4.65 & 3.50  \\
        \bottomrule
    \end{tabular}
    \label{tab:AEC}
\end{table}

\begin{figure*}[ht]
    \centering
    \begin{minipage}{0.24\textwidth}
        \centering
        \includegraphics[width=\linewidth]{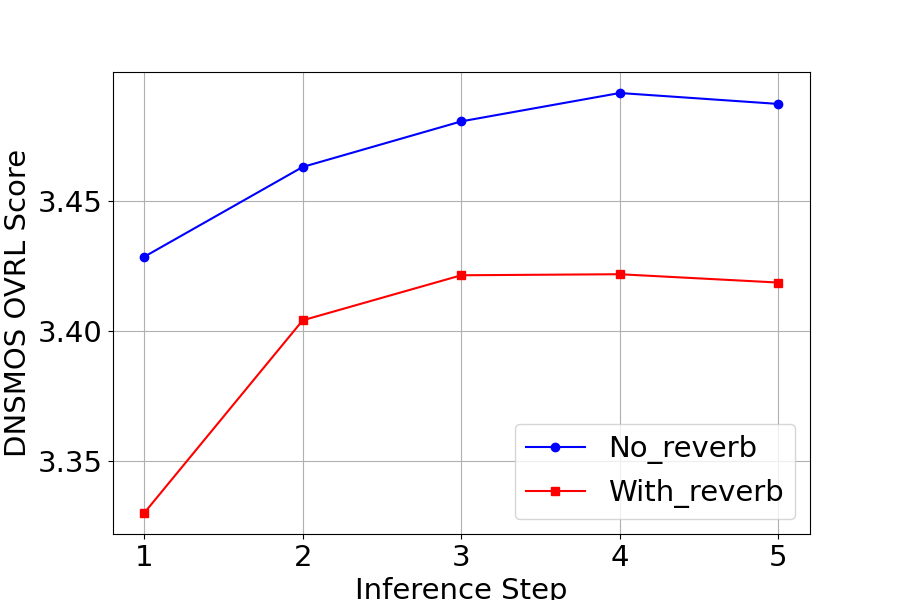}
        \subcaption{NS} \label{fig:b}
    \end{minipage}\hfill
    \begin{minipage}{0.24\textwidth}
        \centering
        \includegraphics[width=\linewidth]{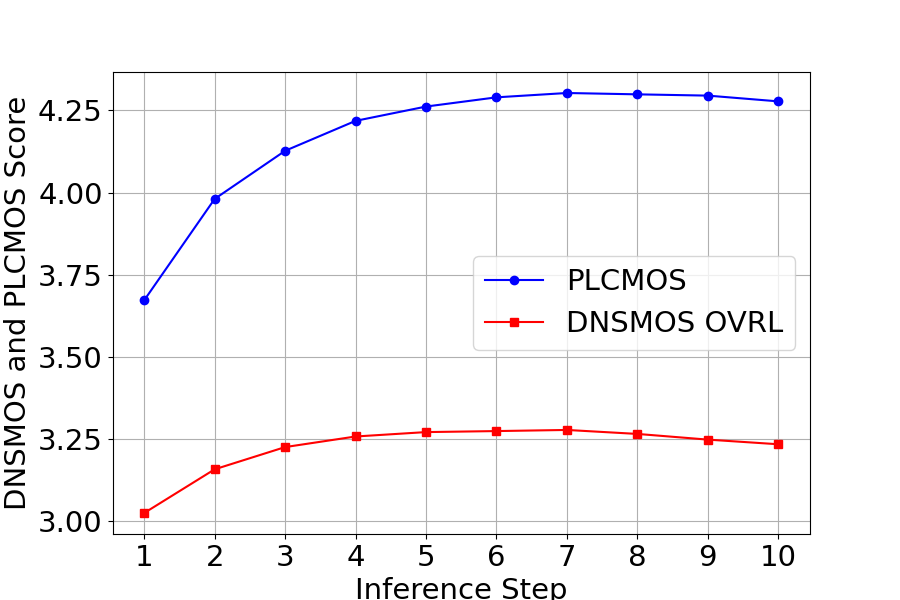}
        \subcaption{PLC} \label{fig:a}
    \end{minipage}\hfill
    \begin{minipage}{0.24\textwidth}
        \centering
        \includegraphics[width=\linewidth]{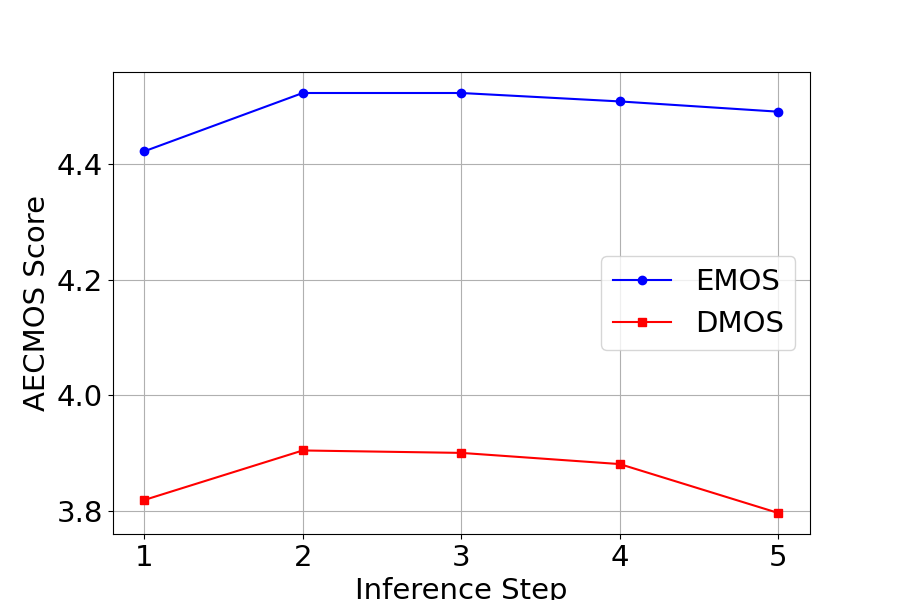}
        \subcaption{AEC} \label{fig:c}
    \end{minipage}\hfill
    \begin{minipage}{0.24\textwidth}
        \centering
        \includegraphics[width=\linewidth]{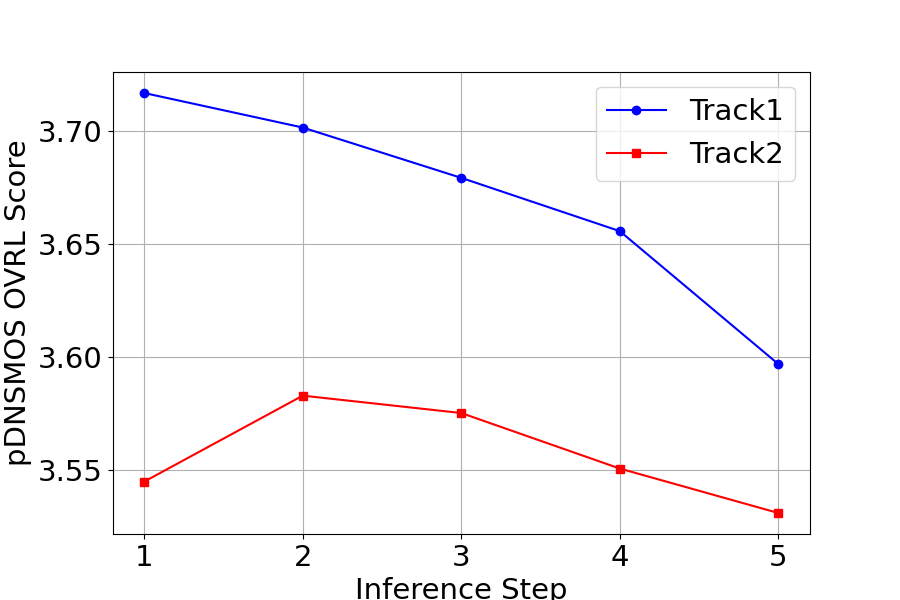}
        \subcaption{TSE} \label{fig:d}
    \end{minipage}
    
    \caption{ Inference-time scaling results on different tasks}
    \label{fig:overall inference}
\end{figure*}

\subsubsection{Emergent Capabilities and Scaling Effects at Test Time}

\textbf{Emergent capabilities.} The SS task is not included in the training data, we use it to test the emergent capabilities of LLaSE-G1.
% We take Libri2mix~\cite{librimix} and WSJ0\_2MIX as our test sets for the speech separation task. We use DNSMOS and SBS as our metrics.
When compared to other discriminative methods, our LLaMA-based LLaSE-G1 demonstrates significant emergent capabilities. With multiple inferences, our generative model outperforms discriminative methods in OVRL scores of 3.17 and 3.25 on test sets, highlighting the potential of LLaSE-G1 to go beyond task-specific optimizations and adapt seamlessly to new tasks.

\begin{table}[ht]
    \centering
    \caption{DNSMOS scores on Libri2mix and WSJ0\_2mix test set.}
    \footnotesize
    \setlength\tabcolsep{1.2pt}
    \begin{tabular}{llcccccc}
        \toprule
        Model & Type & \multicolumn{3}{c}{Libri2mix} & \multicolumn{3}{c}{WSJ0\_2mix} \\
        \cmidrule (lr){3-5} \cmidrule (lr){6-8}
        & & SIG & BAK & OVRL  & SIG & BAK & OVRL \\
        \hline
        Noisy    & -  & 2.33 & 1.66 & 1.64 &  3.42 & 3.20 & 2.76  \\
        Sepformer    & D  & 3.33 & 3.88 & 3.02 &  3.43 & 3.96 & 3.14   \\
        Mossformer2  & D  & 3.44 & 3.94 & 3.11 &  3.50 & 4.05 & 3.23  \\
        \hline
        LLaSE-G1\textsubscript{single} & G  & 3.48 & 3.83 & 3.11 &  3.52 & 3.92 & 3.19  \\
        LLaSE-G1\textsubscript{multi} & G  & \textbf{3.50} & \textbf{3.90} & \textbf{3.17} &  \textbf{3.55} & \textbf{3.97} & \textbf{3.25}  \\
        
        \bottomrule
    \end{tabular}
    \label{tab:Speech Separation}
\end{table}

\textbf{Inference-time scaling.} As shown in Figure~\ref{fig:overall inference}, scaling the inference time improves model performance across nearly all tasks. For the AEC and TSE tasks, performance peaks after the second inference, with EMOS improving from 4.42 to 4.52 and DMOS rising from 3.82 to 3.91. In contrast, the PLC task shows a significant performance boost with increased inference time, with PLCMOS rising from 3.67 to 4.30 and OVRL improving from 3.03 to 3.27, a gain of up to 25\%. For the NS task, the OVRL score increases from 3.42 to 3.49 on the no\_reverb dataset and from 3.33 to 3.42 on the with\_reverb dataset. These results show that scaling test-time compute will initially improve performance, and decrease later due to the accumulation of acoustic distortion.

\subsubsection{Semantic and Speaker Similarity}
As shown in Table~\ref{tab:Similarity}, we compare the semantic and speaker similarity between baseline systems and LLaSE-G1. Notably, TSE and AEC tasks are tested on the blind test sets where ground-truth speech is unavailable. So, we conduct evaluations of NS, PLC, and SS tasks. LLaSE-G1 outperforms generative SE models while getting slightly lower results in SBS, suggesting LLaSE-G1 effectively maintains speech content. Moreover, LLaSE-G1 achieves the highest SimW$_B$ in the NS task and competitive SimW$_B$ in the PLC and SS tasks,  showing superior acoustic preservation capability.

\begin{table}[ht]
    \centering
    \caption{Semantic and speaker similarity results on various tasks, using the same test sets from previous subsections. As for the NS task, we report the average result across the No Reverb and With Reverb test sets.}
    \footnotesize
    \setlength\tabcolsep{4pt}
    \begin{tabular}{lllcc}
        \toprule
        Task & Model          & Type & SBS & SimW$_B$ \\
        \hline
          NS & FRCRN      & D &   \textbf{0.85} &  0.980   \\
             & AnyEnhance & G &   0.82 &  0.970   \\
             & SELM       & G &   0.72 &  0.965   \\
             & GenSE      & G &   0.78 &  0.974   \\
             & LLaSE-G1  & G &   0.83 &  \textbf{0.993}   \\
             % & LLaSE-G1_{multi} & G & 0.80 & 0.988\\
        \hline
        PLC & BS-PLCNet  & D &  \textbf{0.95}   &   \textbf{0.999}  \\
            & LLaSE-G1 & G &  0.85   &  0.992\\
            % & LLaSE-G1_{multi}   & G &  0.74   &   0.980  \\
        \hline
        SS   & Sepformer  & D &  0.85  &   0.980  \\
             & Mossformer2 & D & \textbf{0.87}  &   \textbf{0.991}  \\
             & LLaSE-G1 & G &  0.82  &   0.988  \\
             % & LLaSE-G1_{multi}  & G &  0.80  &  0.985\\
        \bottomrule
    \end{tabular}
    \label{tab:Similarity}
\end{table}

\subsubsection{Subjective Evaluation}
We also conducted evaluations on subjective listening tests and user studies. The experimental setup and results are shown in Table~\ref{tab:subjective_mos}. We primarily used the DNS blind test set for the subjective evaluation. A total of 12 participants were recruited, and each participant was asked to listen to 50 audio samples. The Mean Opinion Score (MOS) was calculated on a 5-point scale, ranging from 1 (poor quality) to 5 (excellent quality). Although GENSE achieves comparable performance to LLaSE-G1 on the DNSMOS metric, LLaSE-G1 still delivers better perceptual quality in subjective listening, further demonstrating its superior capability in preserving acoustic details.

\begin{table}[ht]
    \centering
    \caption{Subjective MOS comparison. “D” denotes discriminative models, “G” denotes generative models.}
    \label{tab:subjective_mos}
    \footnotesize
    \setlength\tabcolsep{6pt}
    \begin{tabular}{lcc}
        \toprule
        Model & Type & MOS \\
        \midrule
        Clean & - & \textbf{4.8125} \\
        FRCRN & D & 3.8917 \\
        SELM & G & 3.7750 \\
        GENSE & G & 4.0083 \\
        LLaSE-G1 & G & 4.1208 \\
        \bottomrule
    \end{tabular}
\end{table}

\subsubsection{Ablation Study}

% We compare DNSMOS scores on the DNS blind test set to evaluate different system configurations started from SELM~\cite{wang2024selmspeechenhancementusing}, focusing on input representation, language model architectures, and decoding systems. Continuous input representation consistently outperforms discrete input with the same Transformer/HiFiGAN~\cite{attentionisallyouneed,HiFiGAN} setup, suggesting it preserves acoustic information better. 
% The X-codec2 decoder shows significant improvements over HiFiGAN, highlighting the effectiveness of modern neural codecs in speech enhancement. Adopting a multi-inference strategy further boosts performance, establishing a new state-of-the-art.

We conduct an ablation study to evaluate the effectiveness of input representations, output representations, and model backbone, choosing SELM as the baseline. As shown in Table~\ref{tab-8}, when replacing inputs and output with proposed continuous features and speech tokens, there is an obvious improvement, revealing the effectiveness of acoustic preservation. Additionally, Experimental results show that while Xcodec2 tokens are designed to better capture acoustic information, discrete tokens still discard more detailed content than continuous embeddings when used as model inputs. Besides, there is no performance drop when replacing the full attention Transformer with casual attention LLaMA layers. Finally, adopting a multi-inference strategy further boosts performance.

\begin{table*}[hbt!]
    \centering
    \caption{DNSMOS scores on DNS blind test set without reverb. "T" represents Discrete tokens, and "E" represents Embeddings. "S" represents Single inference, and "M" represents Multiple inference.}
    \label{tab-8}
    \footnotesize
    \setlength\tabcolsep{1.5pt}
    \begin{tabular}{ll c c c c}
        \toprule
        & Input & LM & Output & Inference & OVRL \\
        \hline
        Noisy & -  & - & -  & - & 2.48  \\
        \hline
        Baseline & WavLM Token & Transformer & HiFiGAN & S & 3.26 \\
        & WavLM Embedding & Transformer & HiFiGAN & S &3.34 \\
        & WavLM Embedding & LLaMA & HiFiGAN & S &3.35 \\
        & Xcodec2 Token & LLaMA & X-codec2 & S &3.29 \\
        Proposed & WavLM Embedding & LLaMA & X-codec2 & S & 3.43 \\
        & WavLM Embedding & LLaMA & X-codec2 & M & \textbf{3.49} \\
        \bottomrule
    \end{tabular}
    \label{tab:Ablation}
\end{table*}

\section{Conclusion}
In this study, we propose LLaSE-G1, a general LLaMA-based framework that unifies a wide range of speech enhancement (SE) tasks. Specifically, we utilize continuous acoustic features as input and predict 1D speech tokens to maximize fidelity to the original audio. To support multiple SE tasks, we design dual-channel inputs and outputs within a unified architecture. Extensive experiments demonstrate that LLaSE-G1 achieves state-of-the-art performance across various benchmarks, establishing it as a strong foundation model. Furthermore, LLaSE-G1 exhibits clear test-time scaling behavior and emerging generalization capabilities to previously unseen SE tasks. Notably, our framework requires no additional prompts to differentiate between SE tasks, instead relying on the LLM’s intrinsic ability to infer the task type from the input itself.

% In this study, we introduce LLaSE-G1, a unified LLaMA-based framework to solve the acoustic preservation and generalization problem of speech enhancement. We use both continuous feature input rather than discrete token input and efficient X-codec2 to maximize acoustic preservation. Additionally, we propose a well-designed multi-task framework that effectively integrates up to five speech enhancement tasks, successfully resolving the input incompatibility between the AEC and TSE tasks. Furthermore, despite being an unseen task during training, the SS task demonstrates outstanding performance, providing strong evidence of our model's emergent capabilities and its potential to tackle additional tasks.

\clearpage
\noindent\textbf{Limitations}

% Although the proposed approach demonstrates promising results, there are several limitations that need to be addressed in future research. Based on pre-trained WavLM and X-codec2, both of which operate at a 16,000 Hz sampling rate, the output of our system is limited to a 16,000 Hz sampling rate as a result. The duration and scaling of the data, as well as the authenticity and diversity of the data augmentation methods used in our system, still need further improvement.

Although LLaSE-G1 demonstrates promising results across diverse SE tasks, there are several limitations that can be addressed towards LLaSE-G2. First, LLaSE-G1 operates at a 16,000 Hz sampling rate due to WavLM and X-codec2. We plan to support full-band audio and super-resolution generation in future research. 
Second, the training data and model size of LLaSE-G1 are relatively small as compared with that used in mainstream audio langauge models for understanding and conversation tasks. Hence we would like to further scale up data and model size to boost performance in generative speech enhancement.
% Second, the training data and model size of LLaSE-G1 are relatively small. Scaling effects at training time can be further explored. 

% \noindent\textbf{Future Work}

% In future work, we aim to address the limitations outlined above and explore several promising directions. First, we plan to extend the system to support higher sampling rates, enabling better preservation of acoustic details and improving overall performance. Second, we will focus on enhancing the duration and scaling of the data, as well as improving the authenticity and diversity of data augmentation methods, to better handle real-world scenarios. Additionally, we will continue to expand our multi-task framework to include more tasks, further strengthen the versatility of our system.

\bibliography{custom}

\clearpage
\appendix

\section{Appendix for Experimental Set Up}
\label{sec:appendix1}

\subsection{Model Configuration}
\label{sec:appendix2}
We use the open-source checkpoints of WavLM-large and X-codec2. The LLaMA-based LM comprises 16 LLaMA layers, each with 16 attention heads, a dropout rate of 0.1, a hidden size of 2048, and an intermediate size of 4096. The total number of parameters in the model is approximately 1.07 billion, which includes all learnable weights and biases across all layers. 
The model has 2 input linear layers and 2 output linear layers. The input layer maps the 1024-dimensional WavLM embedding to another 1024-dimensional vector, while the output layer transforms a 2048-dimensional vector into a 65536-dimensional vector, which is the codebook size of Xcodec2.

We trained the model for 100,000 steps using 4 NVIDIA L40 GPUs, with a batch size of 6 per GPU and the AdamW optimizer. The learning rate is set to 1e-4.

% \subsection{Datasets}
% Clean data consists of DNS challenge datasets~\cite{2020dns,2023dns}, librispeech~\cite{librispeech}, HiFi-TTS~\cite{hifitts}, and subset from AISHELL-1~\cite{aishell1}, AISHELL-3~\cite{aishell3} datasets, which includes English, Chinese, Spanish, German, Italian. All audios will  be resampled to 16,000 Hz.

% Noise data consists of DNS challenge datasets~\cite{2020dns,2023dns}, AEC chanllenge datasets~\cite{2023aec}, DEMAND\footnote{\href{https://www.kaggle.com/datasets/chrisfilo/demand}{DEMAND dataset link}}, ESC-50\footnote{\href{https://github.com/karolpiczak/ESC-50}{ESC-50 dataset link}} sum up to arond 1000 hours. Noise includes animal sounds (e.g., dog barking, bird chirping), natural soundscapes and water sounds (e.g., rain, sea waves), human non-speech sounds (e.g., baby crying, laughter), interior/domestic sounds (e.g., door knocking, washing machine), and exterior/urban noises (e.g., helicopter, car horn)
\subsection{Test Sets}
\textbf{NS:} Interspeech 2020 DNS Challenge blind Test Set.~\cite{2020dns} It contains 600 clips (300 synthetic and 300 real), with synthetic clips generated using clean speech and noise not seen during training, and real clips crowdsourced in diverse noisy conditions.\\
\textbf{PLC:} Interspeech 2022 PLC Challenge test set~\cite{2022plc} This is a realistic evaluation dataset based on packet loss patterns from actual calls, providing a methodology for comparing different approaches and a new objective metric to help researchers improve their techniques.\\
\textbf{TSE:} ICASSP 2023 DNS Challenge blind Test Set~\cite{2023dns} The blind test set includes two tracks Headset and Speakerphone with clips featuring 10-30 seconds of enrollment speech, with or without noise. It is used for final rankings and evaluates both personalized and non-personalized models using the Personalized ITU-T P.835 framework. \\
\textbf{AEC:} ICASSP 2023 AEC Challenge blind Test Set~\cite{2023aec} The blind test set in the AEC Challenge consists of real-world data collected from over 10,000 diverse audio devices and environments. It is used to determine the final competition winners. The dataset includes recordings of both single-talk and double-talk scenarios, with varying conditions like background noise, reverberation, and device distortions. \\
\textbf{SS:} Libri2mix~\cite{librimix}, WSJ0-2mix. These two test sets are commonly used in speech separation, which is mixed from librispeech and WSJ datasets.

\subsection{Baseline Systems}
\textbf{NS:} For discriminative systems, we choose Conv-TasNet~\cite{Luo_2019},DEMUCS ~\cite{défossez2019demucsdeepextractormusic},FRCRN~\cite{zhao2024frcrnboostingfeaturerepresentation} , which is recent SOTA models on  noise suppression. For generative systems, we choose SELM~\cite{wang2024selmspeechenhancementusing}, which introduce LM to speech enhancement, and GenSE~\cite{yao2025gensegenerativespeechenhancement} and AnyEnhance~\cite{zhang2025AnyEnhanceunifiedgenerativemodel}, 2 newly released SOTA-level generative speech enhancement systems.\\
\textbf{PLC:} we use BS-PLCNet ~\cite{zhang2024bsplcnetbandsplitpacketloss}, Team Kuaishow~\cite{2022kuaishou}, which are the winners of the  2024 challenge and 2022 challenge respectively, and other systems in challenge like PLCNet~\cite{liu2022plcnet} and LPCNet~\cite{LPCNet}  as our baseline systems. \\
\textbf{TSE:} We compare our model with two baseline systems: TEA-PSE 3.0~\cite{ju2023teapse30tencentetherealaudiolabpersonalized}, the winner of the challenge, and NAPSE~\cite{yan2023npuelevocpersonalizedspeechenhancement}, which placed second.\\
\textbf{AEC:} For baseline comparison, we choose recent efficient and state-of-the-art systems as our baseline systems, including UCLNet,~\cite{shetu2024ulcnet},AlignUCLNet~\cite{shetu2024alignulcnetlowcomplexityrobustacoustic},AlignCruse~\cite{indenbom2023AlignCruse} and DeepVQE~\cite{indenbom2023deepvqerealtimedeep}, which is the state-of-the-art model in the AEC task.\\
\textbf{SS:} We use SOTA discriminative speech separation systems such as Sepformer~\cite{subakan2021sepformer} and Mossformer2~\cite{zhao2024mossformer2combiningtransformerrnnfree}, which is the SOTA system on speech separation, as our SS baseline systems.\\

\end{document}